\documentclass[letterpaper,10pt]{article}
\usepackage{multicol}
\usepackage{graphicx}
\begin{document}
\begin{center}
{\bf\large Aharonov-Bohm effect and Resonances
	in the circular quantum billiard with two leads}
\vskip 3ex
	{Suhan Ree and L.E. Reichl}\\
	{\em Center for Studies in Statistical Mechanics and Complex Systems}\\ 
	{\em The University of Texas at Austin, Austin, Texas 78712}\\
	(August 31, 1998) 
\vskip 2ex

\parbox{14cm}{
\indent\indent We calculate the conductance through a
circular quantum billiard with two leads and a point magnetic flux at the
center.
The boundary element method is used to solve the Schr\"{o}dinger equation
of the
scattering problem, and the Landauer formula is used to calculate the
conductance
from the transmission coefficients. We use two different shapes of leads,
straight and conic,
and find that the conductance is affected by lead geometry, the relative
positions of the
leads and the magnetic flux.
The Aharonov-Bohm effect can be seen from shifts and splittings of fluctuations.
When the flux is equal to $h/2e$ and the angle between leads
is 180$^\circ$, the conductance tends to be suppressed to zero in the low
energy range
due to the Aharonov-Bohm effect.
\vskip 2ex
\noindent Submitted to Phys.~Rev.~B
}
\end{center}
\vskip 5ex
\begin{multicols}{2}
\section*{I. Introduction}
Aharonov and Bohm, in their historic 1959 paper\cite{ab}, showed theoretically
that charged matter waves are phase shifted if they pass through a
 magnetic vector potential even when there is no magnetic field present in
that region of space. This so-called Aharonov-Bohm effect (AB effect) is a
purely quantum mechanical phenomenon. The magnetic vector potential has no
effect on
classical particles. The AB effect has been observed in several
experiments using electron interferometry\cite{exp}.

The AB effect can also be studied in confined geometries with leads
by looking at behaviors of electron transport through these systems.
An one-dimensional structure such as a small metal ring with a confined magnetic flux
has been used to investigate the connection between the conductance oscillations 
and the AB effect\cite{butt1}-\cite{chan2}.
Recent developments in micrometer-scale 
technology have also made possible new ways to observe the electron transport 
through confined two-dimensional cavities.
These ballistic electron waveguides are formed from the two-dimensional
electron gas (2 DEG) found at the interface 
of GaAs/AlGaAs heterostructure\cite{markus}.
They are formed at very low-temperature and with
sizes of less than $\mu$m to make electrons phase-coherent inside.
Since the quantum mechanics can be used to explain the behavior of electrons,
these devices also provide a natural place to look for 
the AB effect\cite{datta}-\cite{yacoby}.

In these systems, the magnetoresistance has been observed 
to oscillate with a period of $h/e$ 
or $h/2e$ as the magnetic flux is varied. 
In the ballistic regime, 
$h/e$ oscillation is expected due to gauge invariance in the quantum mechanics.
In the diffusive regime 
(when both the elastic mean free path and the phase-coherence
length are not larger than the size of the sample), 
oscillations with period $h/2e$ were observed,
and had been predicted theoretically\cite{aas}.
Kawabata and Nakamura\cite{nakamura} argued that $h/2e$ oscillation is theoretically 
predicted even in ballistic cases
using the semiclassical scattering theory.

In this paper, we study the conductance properties of a two-dimensional
ballistic circular billiard with two leads attached.  
Circular billiards have
been studied before in several contexts.
Lin\cite{lin} calculated the conductance for this system using the semiclassical 
conductance formula.
In Refs.\cite{persson} and \cite{berggren}, 
circular billiards with two leads were investigated
with a homogeneous magnetic field applied.
We will apply a point magnetic flux at the center of the circular  billiard (AB
billiard). Then the cavity contains only the vector potential
with no magnetic field.
In the system we consider here, 
the size of the cavity and the temperature are small enough to make 
the system ballistic,
the boundary is formed by a hard-wall potential, and
electron-electron interaction is ignored.
Under these conditions, the Landauer formula\cite{landauer}-\cite{baranger} 
for two leads 
can be used to obtain the conductance of the system.
(Because $h/e$ oscillation is expected in this ballistic model, we will only look
at the magnetic flux between $-h/2e$ and $h/2e$.)
In addition to considering the effect of the vector potential on
conductance, we will also
consider the effect of the shape of the leads on conductance, and
investigate
conductance for two different shapes of leads, straight and conic (see Fig. 1).
We will find that the shape of the leads and the relative position of leads
can have an important effect on the conductance. We will also find that 
the AB effect can dominate the system under a special circumstance.

In Sec.\ II, we first consider the properties of eigenstates of a closed
circular
billiard with a magnetic flux line down its center. In Sec.\ III,
the open billiard is studied, and the boundary element method which is
used to obtain the conductance is discussed.
In Sec.\ IV, we show our numerical results.
Finally, we summarize in Sec.\ V.

\section*{II. Closed Billiard}
In this section, we obtain the energy levels for an electron
 in a closed circular billiard
with a magnetic flux line down the center. 
Reimann {\it et al}.\cite{reimann} calculated the energy density of states 
semiclassically using the Gutzwiller's trace formula.
In this paper, we must understand this system quantum mechanically 
in order
to properly interpret the behavior of the open billiard with leads attached
in the low energy range.
In Ref.\ \cite{ree}, we studied this closed system when the flux line oscillates
periodically in
time. Then magnetic and electric field are induced in the cavity and the
system can undergo
a transition to chaos. When the flux line is constant in time, only a
magnetic vector potential
appears in the cavity and there is no chaos but we can study the AB effect.

The Hamiltonian of the system can be written
$$
\hat H_0(\hat r,\hat p_r,\hat p_\theta)={\hat p_r^2\over2m}+{(\hat
p_\theta-\alpha\hbar)^2
        \over2m\hat r^2}+\hat V_R(\hat r),
\eqno(1)
$$
where $\alpha=\Phi/\Phi_0$ is a dimensionless quantity when $\Phi$ 
is the magnitude of the magnetic flux
through the center of the cavity and $\Phi_0=h/e$, and $\hat V_R(\hat r)$ is a
hard-wall potential confining
the electron to the interior of a circular billiard
of radius $R$. From here on, we will only consider $\alpha$ in the range
$-0.5<\alpha\le0.5$, which is sufficient because of gauge invariance.
For $\alpha=0$ and $\alpha=0.5$, the energy eigenvalues are two-fold degenerate.
However, the flux breaks the
time-reversal symmetry of the system when $0<|\alpha|<0.5$, and  removes these
 two-fold degeneracies.
 After solving the time-independent Schr\"{o}dinger equation,
$\hat H_0(\alpha)|l,n;\alpha\rangle = E_{ln}(\alpha) |l,n;\alpha\rangle$,
we can find energy eigenstates,
$$
\langle r,\theta|l,n;\alpha\rangle =
J_{|l-\alpha|}\left({\beta_{ln}(\alpha)r\over R}\right)
\ e^{il\theta},
\eqno(2)
$$
and energy eigenvalues,
$$
E_{ln}(\alpha)={\hbar^2 \beta_{ln}(\alpha)^2\over 2 mR^2},\
\eqno(3)
$$
where $m$ is the mass of the electron, and $\beta_{ln}(\alpha)$ is the
$n$th zero of
$J_{|l-\alpha|} (x)$ ($l$: integer, $n$: positive integer).
The energy eigenvalues, $E_{ln}(\alpha)$,  are plotted in Fig.\ 2 as a
function of $\alpha$.
We can easily prove that the energies are symmetric in $\alpha$.
We also see no level-repulsion as expected since this system has no chaos.
When $\alpha=0$, $|l,n;0\rangle$ and $|-l,n;0\rangle$ are
degenerate, and when $\alpha=0.5$, $|l,n;0.5\rangle$ and $|l-1,n;0.5\rangle$
are again degenerate.

\section*{III. Open Billiard}
We now create an open system by attaching two infinite leads to the
circular AB billiard.
As shown in Fig.\ 1, we will study two different lead geometries. For one
geometry,
the leads are straight. For the other geometry, the leads have conic shape.
For both geometries, there are four dimensionless
quantities which characterize the system. There are two geometric
parameters, $\Delta$
(the opening angle of leads) and $\gamma$ (the relative angle
between leads).
There are two physical quantities, $\alpha=\Phi/\Phi_0$
(the dimensionless magnetic flux)
and $\epsilon=(mR^2/\hbar^2)E_F$ (the dimensionless Fermi energy).

For straight leads, we can calculate the number of propagating modes if we
are given
values for $\Delta$ and $\epsilon$. If we define the Fermi wave number as
$k_F=\sqrt{2mE_F}/\hbar$,
there are $n$ propagating modes if
$$
{n\pi\over W}\le k_F < {(n+1)\pi\over W},
\eqno(4)
$$
where $W=2R\,\sin(\Delta/2)$.
In other words, if $\epsilon_n\le\epsilon<\epsilon_{n+1}$, then there are $n$
propagating modes in leads where $\epsilon_n$ is defined as
$$
\epsilon_n={n^2\pi^2\over8\sin^2({1\over2}\Delta)}.
\eqno(5)
$$
All the remaining modes are evanescent modes. Here we are interested in
the energy range,
$\epsilon_0\le\epsilon<\epsilon_2$.

For conic leads, there is no way to distinguish between propagating and
evanescent modes before solving the problem, because all existing modes are
propagating
modes.  But, after solving the equation, we find that, except for a finite
number of
lowest modes, all higher modes will be reflected completely. As we will see
in our
subsequent results,
transmission is not necessarily zero when $\epsilon<\epsilon_1$ for conical
leads,
and the second
propagating mode starts to transmit when $\epsilon$ is lower than $\epsilon_2$.
We will call this ``tunneling." The two lead geometries give similar
results except for this tunneling effect.

We will briefly look at the numerical method. The partial differential equation
we are solving is a Helmholtz equation,
$$
( \nabla^2+k_F^2)\,\Psi(r,\theta)=0, 
\eqno(6)
$$
with the boundary condition, $\Psi|_{\rm on\ all\ boundaries}=0$.
The boundary element method\cite{frohne,na} is used 
to solve this problem. We define a loop $C$
consisting of two circular segments, $P_1$ and $P_2$, and two segments,
$C_1$ and $C_2$
across the
contacts between the cavity and leads (see Fig.\ 1).
For straight leads, $C_1$ and $C_2$ are linear segments, but
for conic leads they are parts of a circle, making $C$ a complete circle.
The loop $C$ divides the whole region into three parts consisting of the
interior
of the circular cavity, the lead I, and the lead II.
We will denote basis states for the interior of the circular cavity by
$\{\phi^{(c)}_l|l:{\rm integer}\}$;
for lead I, $\{{\phi^{(1)}_{n_1}}^\pm | {n_1}\ge1\}$; and for lead II,
$\{{\phi^{(1)}_{n_2}}^\pm | {n_2}\ge1\}$;
where + ($-$) stands for an outgoing (incoming) mode.
All basis states satisfy the Helmholtz equation.
For example,
$$
( \nabla^2+k_F^2)\,\phi^{(c)}_l(r,\theta)=0,
\eqno(7)
$$
and so on.
For both lead geometries, $\phi^{(c)}_l$ is given by
$$
\phi^{(c)}_l(r,\theta)=c^{(c)}_l\,J_{|l-\alpha|}(k_Fr)\ e^{il\theta}.
\eqno(8)
$$
For the straight leads, we have
$$
{\phi^{(i)}_{n_i}}^\pm(x_i,y_i)=\cases{c^{(i)}_{n_i}\,\sin({n_i}\pi y_i/ W)\,
                                        e^{\pm ik_{n_i}x_i} &
                                                $(1\le {n_i}\le N_p)$;\cr
                                   c^{(i)}_{n_i}\,\sin({n_i}\pi y_i/ W)\,
                                        e^{-\kappa_{n_i}x_i} &
                                                $(N_p<{n_i})$,\cr}
\eqno(9)
$$
where $(x_i,y_i)$'s are local coordinates,
$k_{n_i}=\sqrt{(2m/\hbar^2)E_F-
({n_i}^2\pi^2/ W)^2}$, and \\
$\kappa_{n_i}=\sqrt{({n_i}^2\pi^2/ W)^2-
(2m/\hbar^2)E_F}$.
(In this case, $\alpha$ is set to zero inside the leads 
as an approximation to simplify the bases. $i=1,2$.)
We have $N_p$ propagating modes where $N_p$ is the largest integer $n$
satisfying
$(n^2\pi^2/ W)^2\le (2m/\hbar^2)E_F$.

For the conic leads, we have
$$
{\phi^{(i)}_{n_i}}^+(r_i,\theta_i)=c^{(i)}_{n_i}\,H^{(2)}_{{n_i}\pi/\Delta}(
k_Fr_i)\,
        \sin\left({{n_i}\pi\theta_i\over \Delta}\right)\,e^{i\alpha\theta_i},
\eqno(10)
$$
and
$$
{\phi^{(i)}_{n_i}}^-(r_i,\theta_i)=c^{(i)}_{n_i}\,H^{(1)}_{{n_i}\pi/\Delta}(
k_Fr_i)\,
        \sin\left({{n_i}\pi\theta_i\over \Delta}\right)\,e^{i\alpha\theta_i},
\eqno(11)
$$
where $(r_i,\theta_i)$'s are local coordinates and $H^{(1)}_\nu$ and
$H^{(2)}_\nu$ are the Hankel functions ($i=1,2$).
The normalization constants, $c^{(i)}_{n_i}$'s, are
chosen to make the incoming flux of electrons unity.

The solution to the Helmholtz equation, Eq.\ (6), can be expressed as follows.
Inside the cavity,
$$
\Psi^{(c)}(r,\theta)=\sum^\infty_{l=-\infty}\, a_l\phi^{(c)}_l(r,\theta);
\eqno(12)
$$
in lead I with an incoming wave of $k$th mode,
$$
\Psi^{(1)}(r_1,\theta_1)={\phi^{(1)}_k}^-(r_1,\theta_1)
                +\sum^\infty_{j=1}\, r_{jk}{\phi^{(1)}_j}^+(r_1,\theta_1);
\eqno(13)
$$
and in lead II,
$$
\Psi^{(2)}(r_2,\theta_2)=\sum^\infty_{j=1}\,
t_{jk}{\phi^{(2)}_j}^+(r_2,\theta_2),
\eqno(14)
$$
where $r_{jk}$ and $t_{jk}$ are reflection and transmission amplitudes,
respectively.

Let us now consider the boundary conditions on loop $C$. (The boundary
conditions on the walls
of leads are already taken care of, because the bases we chose in leads
satisfy those
conditions.)
On boundaries $C_1$ and $C_2$, we have the boundary conditions,
$$
\Psi^{(c)}|_{{\rm on}\ C_i}= \Psi^{(i)}|_{{\rm on}\ C_i},
\eqno(15)
$$
and
$$
\left.{\partial\Psi^{(c)}\over\partial n}\right|_{{\rm on}\ C_i}=
        \left.{\partial\Psi^{(i)}\over\partial n}\right|_{{\rm on}\ C_i},
\eqno(16)
$$
where $n$ is a normal coordinate of boundaries ($i=1,2$).
To express the boundary condition on $P_1$ and $P_2$, we will introduce
bases, $\{\xi^{(1)}_{\nu_1}|\nu_1:{\rm integer}\}$ and
$\{\xi^{(2)}_{\nu_2}|\nu_2:{\rm integer}\}$.
They are defined on $P_i$ as
$$
\xi^{(i)}_{\nu_i}(\lambda_i)={1\over\sqrt{\Delta_i}}\,e^{i{2\nu_i\pi\over
\Delta_i}\lambda_i},
\eqno(17)
$$
where $\lambda_i$ is a longitudinal coordinate along the boundary $P_i$ and
$\Delta_i$ is the
size of the boundary $P_i$ ($i=1,2$). Then these boundary conditions are
written as
$$
\Psi^{(c)}|_{{\rm on}\ P_i}= 0,
\eqno(18)
$$
and
$$
\left.{\partial\Psi^{(c)}\over\partial n}\right|_{{\rm on}\ P_i}=
        \sum^\infty_{\nu_i=-\infty}\, b^{(i)}_{\nu_i} \xi^{(i)}_{\nu_i}.
\eqno(19)
$$

We multiply  Eq.\ (7) by $\Psi^{(c)}$ and we multiply Eq.\ (6) by
$\phi^{(c)}_l$,
and subtract one
from another.  Then, we integrate over the loop $C$. From the Green's
theorem, we get the
equation,
$$
\oint_C\,d\lambda\, \left[\Psi^{(c)}{\partial\phi^{(c)}_l\over\partial n}-
                {\partial\Psi^{(c)}\over\partial n}\phi^{(c)}_l\right]=0.
\eqno(20)
$$
$\Psi^{(c)}$ and ${\partial\Psi^{(c)}\over\partial n}$ are obtained
using Eqs.\ (15), (16), (18), and (19).
With suitable truncations, after integrating over
each segment, we finally obtain a matrix equation ${\bf A}\cdot{\bf x}={\bf B}$.
Solving this matrix equation gives us the values for $r_{ij}$, $t_{ij}$,
$b^{(1)}_{\nu_1}$, and $b^{(2)}_{\nu_2}$. ($a_l$'s can be obtained using these
values following similar steps.)

The reflection and transmission amplitudes, $r_{ij}$ ($i,j\le N_p$)
and $t_{ij}$ ($i,j\le N_p$), respectively,
determine the
left half of the {\bf S} matrix,
$$
{\bf S}=\pmatrix{{\bf r} & {\bf t'} \cr {\bf t} & {\bf r'}\cr},
\eqno(21)
$$
 for the open circular billiard.
In Eq.\ (21), {\bf r}, {\bf t}, ${\bf r'}$, and ${\bf t'}$ are $(N_p\times
N_p)$-submatrices
of {\bf S}, and  $N_p$ is
the number of channels  in the lead.
The quantities {\bf r} and {\bf t} are matrices
containing
the reflection amplitudes and the
transmission amplitudes,
respectively, when there is an incoming wave in lead I.
Also, the right half of the {\bf S} matrix, ${\bf r'}$ and ${\bf t'}$, 
can be obtained by having an incoming wave
in the lead II instead of the lead I. But, in our geometries, it can be done
by changing the direction of the magnetic flux.

The conductance for this
system is obtained from the transmission amplitudes, $t_{ij}$,
using the Landauer formula.
For transmission from lead I to lead II
the Landauer formula is given by
$$
G_{I\rightarrow II}={2e^2\over h} \sum^{N_p}_{i,j=1} |t_{ij}|^2.
\eqno(22)
$$
On the other hand, the conductance from lead II to lead I is given by
$G_{II\rightarrow I}=(2e^2/ h) \sum^{N_p}_{i,j=1} |t'_{ij}|^2$.
But, in two-lead conductors, $G_{I\rightarrow II}$ should be 
equal to $G_{II\rightarrow I}$ due to the reciprocity of {\bf S}
matrix\cite{butt2}.

\section*{IV. Numerical Results}
Let us now show some of our results.  The value of $\Delta$ is set to
$20^\circ$ throughout, $\gamma$ is $180^\circ$ or $90^\circ$.
Here
$\epsilon$ is varied from $\epsilon_0$ to $\epsilon_2$ where
$\epsilon_0=0$, $\epsilon_1
=40.91$, and $\epsilon_2=163.7$.
The conductance is calculated using the Landauer formula, Eq.\ (22). 
We define the transmission probability
$T_i\equiv\sum^{N_p}_{j=1}|t_{ji}|^2$  ($i\le N_p$) when there is an $i$th 
incoming wave in lead I. For each $i$, $T_i+R_i=1$ is satisfied due to 
the unitarity of the {\bf S} matrix when $R_i=\sum^{N_p}_{j=1}|r_{ji}|^2$.
Then the conductance
from lead I to lead II is
$$
G_{I\rightarrow II}={2e^2\over h}\, \sum^{N_p}_{i=1} T_i.
\eqno(23)
$$
We will compute $T_1$ and $T_2$ for $\epsilon_0<\epsilon<\epsilon_2$ for
the two different lead geometries. 

Figure 3 shows our results for the conductance for the case of straight
leads for
$\epsilon_0<\epsilon<\epsilon_2$ when $\gamma=180^\circ,
90^\circ$ and $\alpha=0,0.25$. Only $T_1$ is plotted versus $\epsilon$.
Note that for straight leads, $T_1$ is zero for $\epsilon<\epsilon_1$, and
$T_2$ is zero for
$\epsilon<\epsilon_2$.

Figure 4 shows the analogous results for the case of conic leads for
$\epsilon_0<\epsilon<
\epsilon_2$ when $\gamma=180^\circ, 90^\circ$ and $\alpha=0,0.25$.
We see that $T_1$ ($T_2$) starts to emerge
when $\epsilon$ is lower than $\epsilon_1$ ($\epsilon_2$).
It is useful to note that this case has clear computational advantages. When we integrate over
$C_i$'s, the Bessel
functions are constant, while it is not true in the case of straight leads. As
a result, the computation was much faster. Furthermore, we can keep
$\alpha{\neq}0$ also
in leads by choosing suitable bases in leads without spending more
significant computer
time.

There are several conclusions we can draw from these results.
First, the shape of the lead matters. For the conical leads, the particle
can tunnel through
the circular cavity at energies which forbid any propagation in the
straight leads.
On the other hand, away from tunneling energies,
both lead geometries give us similar results in the most of the range we
looked at.

Second, we can see patterns of fluctuations in all cases in Figs.\ 3 and 4.
They have peaks and valleys. (Most of valleys go down to zero. We will call them
``transmission zeros".
We believe that valleys (the transmission zeros, in particular) are related
to energy eigenvalues of the closed billiard.
Although there is no exact one-to-one correspondence, we see loose connections,
especially in Figs.\ 3(b) and 4(b).
Therefore, these transmission zeros (which are reflection peaks)
can be treated as resonances with the cavity.

Third, our results depend on the relative position of the leads. Although the
patterns look similar,
we can see that the locations of peaks and valleys are different for the case
$\gamma=180^\circ$ and the case $\gamma=90^\circ$. Transmission
zeros for the case
 $\gamma=90^\circ$ are more likely to match with energy eigenvalues of
the closed
billiard (see Figs.\ 3(b) and 4(b)). This is reasonable because
when $\gamma\ne180^\circ$,
electrons are more
likely to feel the circular billiard.

Fourth, we can see an indirect result of the AB effect. When $\alpha=0.25$,
we see almost twice as many fluctuations as when $\alpha=0$. 
This means that, if we increase
$\alpha$ from zero, fluctuations will split and be shifted at the same time.
This is due to the AB effect.

Let us now look at the conductance from another point of view. In Fig.\ 5,
we computed $G_{I\rightarrow II}$
of the case of conic leads assuming the energy is a complex number.
We see a series of poles near the real axis.
Such poles have been seen for simple systems relating to transmission zeros
\cite{na},\cite{bagwell},\cite{shao}.
However, we cannot establish an
exact correspondence because our conductance curves are too crowded with
fluctuations.
One interesting feature, however,
is the big poles which follow a line
across the  complex plane when $\gamma=180^\circ$ (see Figs.\ 5(a) and 5(c)).
When $\gamma=90^\circ$, 
Figs.\ 5(b) and 5(d) show that they become less regularly distributed.

Let us now show our results for the special case of $\alpha=0.5$.
In Fig.\ 6, we show graphs of $T_1$ and $T_2$ as a function of $\epsilon$ 
for the cases of conic leads
when $\gamma=180^\circ$ (Fig.\ 6(a)) and $\gamma=90^\circ$ (Fig.\
6(b)).
The graphs for the cases of straight leads show almost the same results, but are
not shown here.

In Fig.\ 6(b), when $\gamma=90^\circ$, we see the same type of
fluctuation pattern that we saw in Figs.\ 3 and 4.
In Fig.\ 6(a), however, when $\gamma=180^\circ$, we see the shut-down of
the conductance
in the lower energy range. This can be
roughly explained using concepts from the theory of path integrals.
The transmission amplitude from lead I to lead II is approximately
proportional to the Green's function of Eq.\ (6)
from the point at the center of $C_1$ to the
point
at the center of $C_2$ when $\Delta$ is small. 
The propagator can be obtained using the
path integral
by summing over all paths. These paths can be divided into two groups,
upper and lower paths. 
Since this particular case has the reflection symmetry 
along an axis parallel to leads,
each lower path is the exact mirror image of an upper path, and they
cancel each other due to the AB effect only when $\alpha=0.5$. 
This is destructive quantum interference.
If the propagator vanishes, the Green's function vanishes, too.
Therefore, the conductance almost vanishes in this case.
This is true only when the Fermi wavelength of the electron is comparable to
the width of the opening in the lower energy range.  As energy gets bigger,
this AB effect will be less pronounced because of the finiteness of the 
width of leads.
\section*{V. Conclusions}
In this idealized model, we have calculated the coonductance through the
circular AB billiard with two leads using the Landauer formula.
We have shown that the shape and the relative position of leads can play
important roles in transport properties of the open AB billiard. We have
observed generic fluctuation patterns, which can be interpreted as resonances.
Transmission zeros tend to correspond well with energy eigenvalues of the closed billiard,
especially when $\gamma=90^\circ$.  
We have also calculated the conductance for complex energies, and found regularly 
located big poles.
We have seen the significant AB effect
in the special case of $\gamma=180^\circ$ and $\alpha=0.5$. Then the
conductance
was suppressed to zero in the lower part of the energy range we looked at.
\section*{Acknowledgments}
The authors wish to thank the Welch Foundation, Grant
No.F-1051, NSF Grant No.INT-9602971, and DOE contract No.DE-FG03-94ER14405
for partial support of this work.
We thank NPACI and the University of Texas High Performance Computing Center for use
of its
facilities.
The authors also wish to thank Kyungsun Na and Professor German
Luna-Acosta for many
helpful discussions, and Professor Alex de Lozanne for helpful discussions
concerning experiments.


\end{multicols}
\twocolumn
\begin{figure}
 \begin{center}
	 \scalebox{0.45}[0.45]{\includegraphics{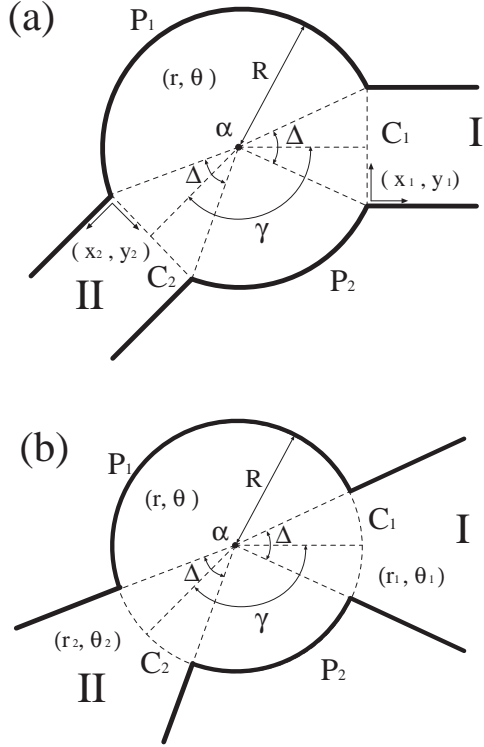}}
 \end{center}
	\caption{
		Geometries of two models. (a) Circular AB billiard 
		with two straight leads. (b) Circular AB billiard with two conic
		leads. In both cases, the magnetic point flux $\alpha$ is placed at the 
		center. $R$ is the radius of the billiard, $\gamma$ is the angle 
		between two leads, and $\Delta$ is the opening angle of leads.
		}
\end{figure}
\begin{figure}
 \begin{center}
	 \scalebox{0.45}[0.45]{\includegraphics{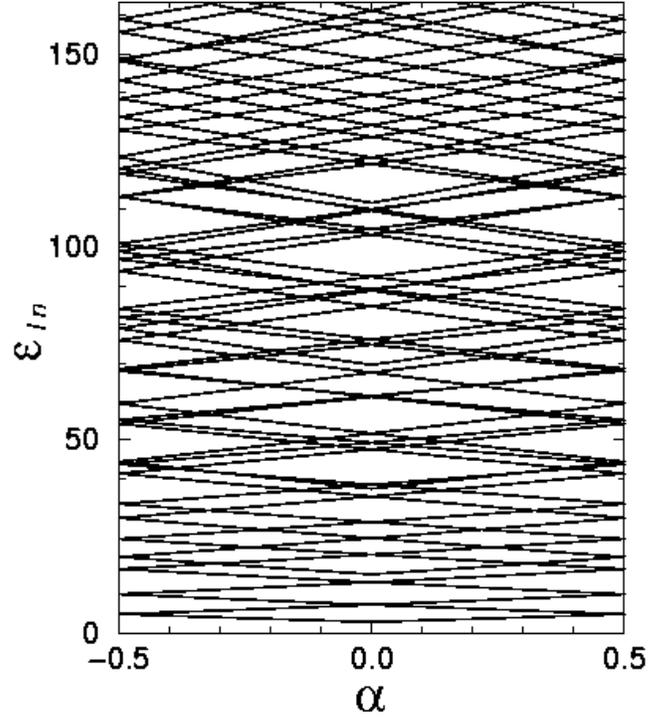}}
 \end{center}
	\caption{
		Energy eigenvalues of the closed circular AB billiard
		as a function of the magnetic flux $\alpha(=\Phi/\Phi_0)$. 
		$\epsilon_{ln}$ is a dimensionless quantity
		defined by $\epsilon_{ln}=(mR^2/\hbar^2)\,E_{ln}$.
		}
\end{figure}
\begin{figure}
 \begin{center}
	 \scalebox{0.85}[0.85]{\includegraphics{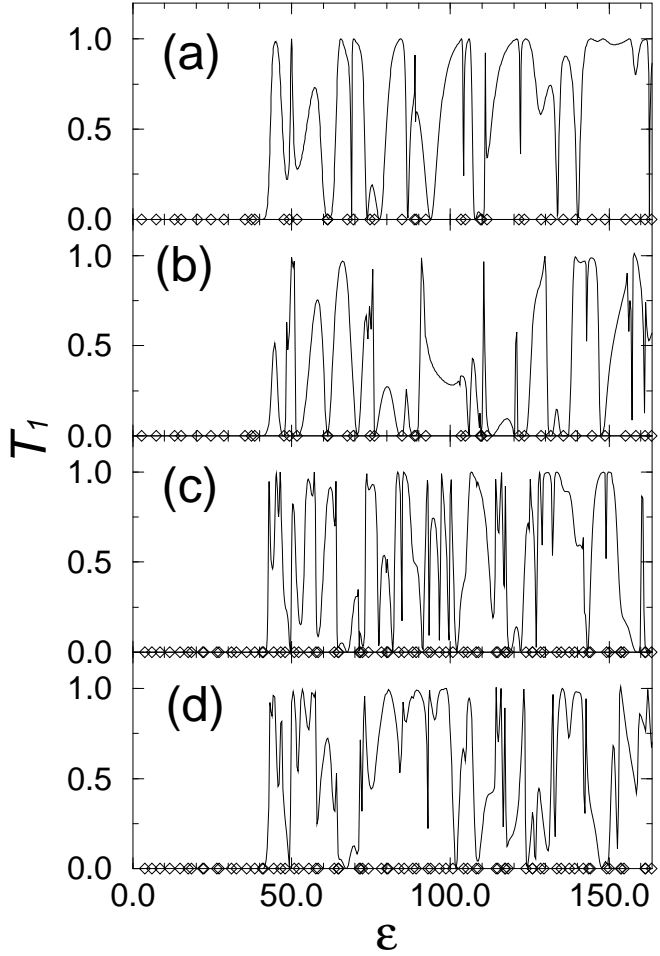}}
 \end{center}
	\caption{
		$T_1$ (transmission probability for the first mode incoming) 
		vs. Fermi energy $\epsilon$ for the case of straight
		leads when $\Delta=20^\circ$. 
		(a) $\alpha=0$, $\gamma=180^\circ$. 
		(b) $\alpha=0$, $\gamma=90^\circ$.
		(c) $\alpha=0.25$, $\gamma=180^\circ$. 
		(d) $\alpha=0.25$, $\gamma=90^\circ$.
		Points($\diamond$) show the energy eigenvalues of the closed AB billiard with
		corresponding $\alpha$ value.
		$\epsilon=(mR^2/\hbar^2) E_F$. $\epsilon$ is in the range of $\epsilon_0
		<\epsilon<\epsilon_2$ where $\epsilon_0=0$, $\epsilon_1=40.91$, and 
		$\epsilon_2=163.7$. $T_i(i>1)$ is equal to zero in this range. 
		The conductance $G_{I\rightarrow II}$ is obtained from $(2e^2/h)T_1$.
		}
\end{figure}
\begin{figure}
 \begin{center}
	 \scalebox{0.85}[0.85]{\includegraphics{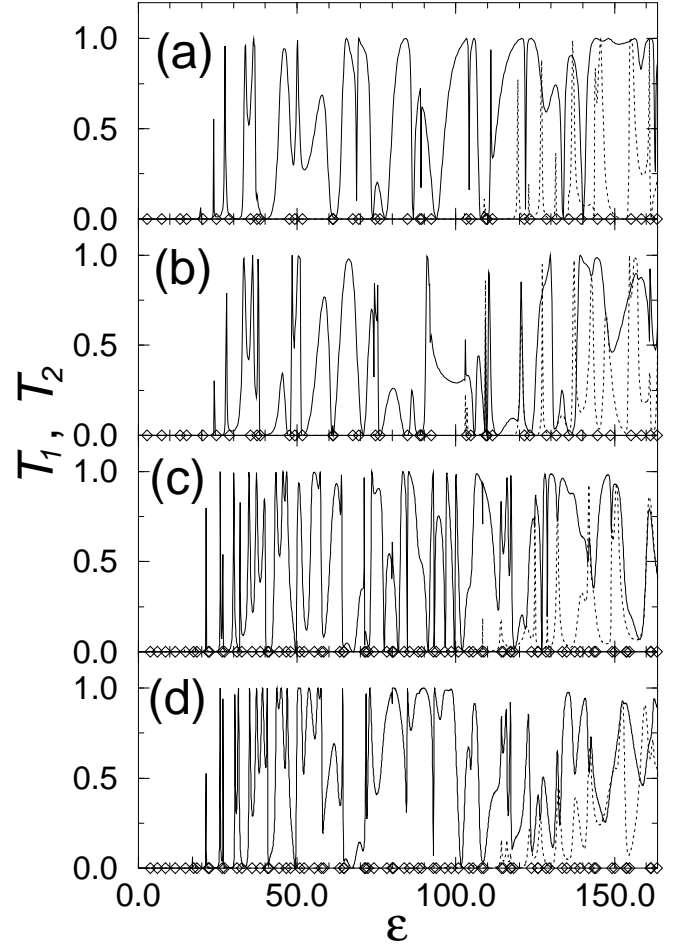}}
 \end{center}
	\caption{
		$T_1$ and $T_2$ (transmission probability for the first or second mode incoming,
		respectively)
		vs. Fermi energy $\epsilon$ for the case of conic
		leads when $\Delta=20^\circ$. 
		(a) $\alpha=0$, $\gamma=180^\circ$. 
		(b) $\alpha=0$, $\gamma=90^\circ$.
		(c) $\alpha=0.25$, $\gamma=180^\circ$. 
		(d) $\alpha=0.25$, $\gamma=90^\circ$.
		Points($\diamond$) show the energy eigenvalues of the closed AB billiard with
		corresponding $\alpha$ value.
		$T_1$ (solid) starts to emerge before $\epsilon_1$, and $T_2$ (dotted) 
		emerges before $\epsilon_2$.
		The conductance $G_{I\rightarrow II}$ is obtained from $(2e^2/h) (T_1+T_2)$.
		}
\end{figure}
\begin{figure}
 \begin{center}
	 \scalebox{0.45}[0.45]{\includegraphics{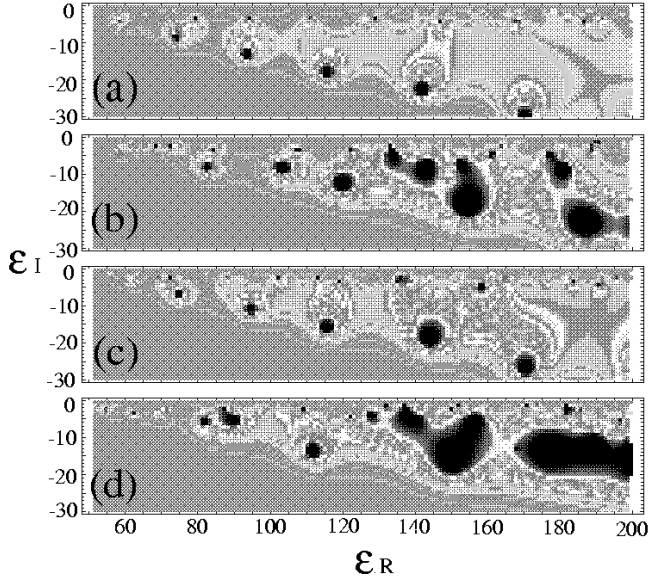}}
 \end{center}
	\caption{
		Conductance $G_{I\rightarrow II}$ vs. complex energy $\epsilon_R+i\epsilon_I$
		for the case of conic leads when $\Delta=20^\circ$.
		(a) $\alpha=0$, $\gamma=180^\circ$. 
		(b) $\alpha=0$, $\gamma=90^\circ$.
		(c) $\alpha=0.25$, $\gamma=180^\circ$. 
		(d) $\alpha=0.25$, $\gamma=90^\circ$.
		Dark spots represent poles in the complex plane. Small poles near the real
		axis ($\epsilon_I=0$) correspond to fluctuations we saw in Fig.\ 4. But, 
		we also observe big poles marching down regularly.
		}
\end{figure}
\begin{figure}
 \begin{center}
	 \scalebox{0.85}[0.85]{\includegraphics{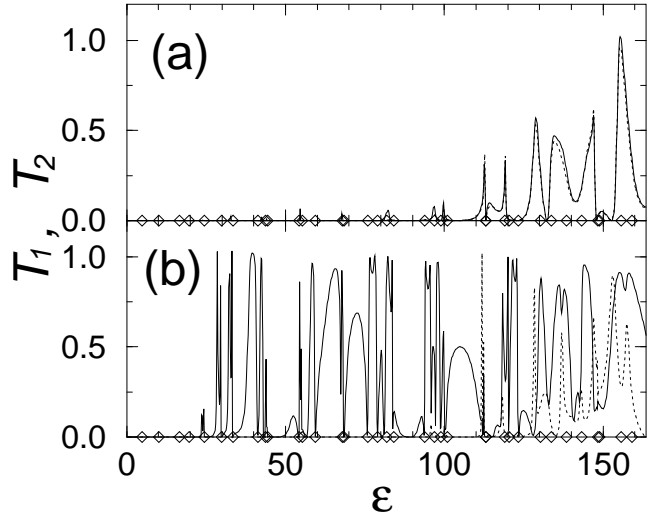}}
 \end{center}
	\caption{
		$T_1$ and $T_2$ vs. Fermi energy $\epsilon$ for the case of conic
		leads when $\Delta=20^\circ$ and $\alpha=0.5$. 
		(a) $\gamma=180^\circ$. 
		(b) $\gamma=90^\circ$.
		Points($\diamond$) show the energy eigenvalues of the closed AB billiard with
		$\alpha=0.5$.
		In (a), $T_1$ (solid) and $T_2$ (dotted) are suppressed to zero in the 
		lower region due to the Aharonov-Bohm effect. But, in (b), 
		fluctuations are observed as in Fig.\ 4.
		}
\end{figure}
\end{document}